\documentstyle[preprint,revtex]{aps}
\begin{document}
\begin{title}
{\bf SO(10) COSMIC STRINGS AND \\
BARYON NUMBER VIOLATION}
\footnote{This work is supported in part by funds provided
by the U. S. Department of Energy (D.O.E.) under
contract \#DE-AC02-76ER03069}
\end{title}
\author{Chung-Pei Ma}
\begin{instit}
Center for Theoretical Physics\\
Laboratory for Nuclear Science
and Department of Physics\\
Massachusetts Institute of Technology\\
Cambridge, Massachusetts 02139 U.S.A.
\end{instit}

\begin{abstract}
SO(10) cosmic strings formed during the phase transition
Spin(10) $\rightarrow$ SU(5) $\times{\cal Z}_2$ are studied.
Two types of strings --- one effectively Abelian and one non-Abelian
--- are constructed and the string solutions are calculated numerically.
The non-Abelian string can catalyze baryon number violation via
the ``twisting'' of the scalar field which causes mixing of leptons
and quarks in the fermion multiplet.  The non-Abelian string is
also found to have the lower energy possibly for the entire range of
the parameters in the theory.  Scattering of fermions in the fields of
the strings is analyzed, and the baryon number violation cross section
is calculated.  The role of the self-adjoint parameters is discussed and
the values are computed.
\end{abstract}
\narrowtext

\newpage
\section{INTRODUCTION}
Motivated by the Callan-Rubakov effect in the context of magnetic
monopoles \cite{callan}, studies have been carried out recently
on the possibility that cosmic strings can also catalyze
baryon-number violation with strongly enhanced cross sections.
It has been shown that the wave function of a fermion scattering
off a cosmic string can acquire a large amplification factor near
the core of the string, leading to enhancement of the processes
that violate baryon number inside the string \cite{alford,perkin}.
The catalysis processes that have been studied include those
mediated by scalar fields and by the grand-unified X and Y gauge
bosons in the string core.  Although strings, in contrast to
monopoles, have no magnetic fields outside, fermions can interact
quantum-mechanically with the long-range gauge fields via the
Aharonov-Bohm effect.  Depending on the flux of the string and
the core model used, the enhanced catalysis cross sections (per
length) can be of the scale of strong interactions in comparison to
the much smaller geometrical cross section $\sim \Lambda_{GUT}^{-1}$,
where $\Lambda_{GUT} \sim 10^{16}$ GeV.  In the early universe when
the density of cosmic strings is high, such processes can play
important roles, washing out any primordially-generated baryon
asymmetry \cite{RB1}, or conceivably even generating the baryon to
entropy ratio observed today.

Cosmic strings can be produced during certain phase transitions
when a gauge group G is broken down to a subgroup H by the vacuum
expectation value of some scalar field $\phi$.  The topological
criterion for the existence of a string is a nontrivial fundamental
homotopy group of the vacuum manifold G/H, denoted by
$\pi_1(\hbox{G}/\hbox{H})$.  For a connected and simply-connected
G, the general construction of the scalar field at large distances
from the string is given by
\begin{equation}
        \phi(\theta) = g(\theta) \phi_0\,,
        \quad g(\theta) = e^{i\tau\theta}\,.
\end{equation}
Here $\tau$ is some generator of G, $\theta$ is the azimuthal
angle measured around the string, and $g(0)$ and $g(2\pi)$
belong to two disconnected pieces of H.  In the papers
referenced in the previous paragraph, the scalar field
responsible for the formation of the string is taken to
have the simple form $\phi(\theta) = e^{i\tau\theta} \phi_0
= e^{i\theta} \phi_0$.  As a result, a non-Abelian string can be
modeled by a U(1) vortex, and the scattering of fermions in the
background fields of the string is governed by the Abelian Dirac
equation.  In general however, for a given $\phi_0$, the generator
$\tau$ can be chosen such that $e^{i\tau\theta} \phi_0$ ``twists''
around the string in more complicated fashion than a phase
$e^{i\theta}$ times $\phi_0$.  This gives rise to dynamically
different strings which are intrinsically non-Abelian
\cite{leandros}.  One expects the complexity and rich
structure of such strings to lead to interesting effects
on fermions traveling around them.  In particular, we will
demonstrate in this paper that for certain $\tau$'s, the
twisting of $\phi(\theta)$ can result in mixing of lepton
and quark fields, providing a mechanism for baryon number
violations distinct from the processes in Abelian strings
studied previously.

Since no strings are formed in the minimal SU(5) model, we choose
the gauge group SO(10) \cite{so10} in this paper as an example of
grand unified theories in investigating the B-violating process.
We will construct string configurations, solve numerically for the
undetermined functions, and study the baryon catalysis in the SO(10)
theory, although we expect such processes to occur in other
non-Abelian theories as well.  In SO(10), stable strings can
be formed when Spin(10) --- the simply-connected covering group
of SO(10) --- is broken down to SU(5)$\times {\cal Z}_2$ by the
vacuum expectation value of a Higgs field $\phi$ in the {\bf 126}
representation \cite{kibble}.  The generators of SO(10) transform
as the adjoint {\bf 45}, which transforms as {\bf 24} + {\bf 1}
+ {\bf 10} + $\bf{\bar{10}}$ under SU(5).  The {\bf 24} and {\bf 1}
generate the subgroup SU(5)$\times$U(1), where the U(1) includes
simultaneous rotations in the 1-2, 3-4, 5-6, 7-8, and 9-10 planes.
We are interested in the generators outside SU(5) because to have
noncontractible loops at all, $g(\theta)$ in Eq.~(1) has to be
outside the unbroken H for some $\theta$.  We will refer to the
U(1) generator as $\tau_{\rm all}$ and to any of the other 20
basis generators outside SU(5) as $\tau_1$; we name the
associated strings as string-$\tau_{\rm all}$ and string-$\tau_1$,
respectively.  As we shall see, the scalar field of string-$\tau_1$
causes mixing of leptons and quarks while string-$\tau_{\rm all}$
is effectively Abelian and no such mixing occurs.  Properties of
string-$\tau_{\rm all}$ such as the string mass per unit length
\cite{everett} and its superconducting capability in terms of
fermion zero modes \cite{witten} have been studied.  We will
compare it with string-$\tau_1$, which will be the main subject
of study of this paper.

In Sec.~II, we give more detailed discussion of the Higgs {\bf 126}
and the breaking of Spin(10) to SU(5)$\times {\cal Z}_2$, and
elaborate on the B-violating mechanism due to the nontrivial
winding of the Higgs field.  In Sec.~III, we write down an
{\it ansatz\ } for the field configuration of each string and
derive the corresponding equations of motion. The numerical
solutions and the energy of the strings are presented in Sec.~IV,
where we find that $\tau_1$-strings have lower energy than
$\tau_{\rm all}$-strings, probably for the entire range of the
parameters in the theory.  Having shown that such strings are
energetically favorable, we turn to the scattering problem in
Sec.~V, where the Dirac equation in the background fields of
the strings is solved, and the differential cross section for
the B-violating processes in string-$\tau_1$ is calculated.
We also comment on the role of the self-adjoint parameters and
compute their values using our string solutions.  To establish
a common notation and to facilitate reading of this paper,
we include in the Appendix a discussion about the relevant
aspects of the spinor representation {\bf 16} of SO(10), which
accommodates a single generation of left-handed fermions.

\section{SO(10) strings}
There is considerable freedom in the breakings of SO(10) down
to the low energy gauge group SU(3)$\times$U(1).  Two commonly
studied examples include the breaking via an intermediate SU(5),
SO(10)$\rightarrow$SU(5), and the one via an intermediate
Pati-Salam SU(4)$\times$SU(2)$_L\times$SU(2)$_R$ \cite{pati}.
Details of the symmetry breaking patterns and the Higgs fields
inducing the breakings can be found in Ref.~6 and the papers
by Slansky and Rajpoot \cite{slansky}.  Kibble, Lazarides and
Shafi argued that the strings formed during the phase transition
SO(10) $\rightarrow$SU(4)$\times$SU(2)$_L\times$SU(2)$_R$ become
boundaries of domain walls \cite{kibble}.  Thus in this paper we
choose the SU(5) breaking pattern instead for its simplicity.
More precisely, we study strings formed when
Spin(10)$\rightarrow$SU(5)$\times{\cal Z}_2$ by the vacuum
expectation value of a Higgs {\bf 126} $\phi$.  The nontrivial
element of ${\cal Z}_2$ corresponds to rotation by 2$\pi$ in SO(10).
The homotopy group $\pi_1(\hbox{Spin(10)}/\hbox{SU(5)}\times
{\cal Z}_2)$ is ${\cal Z}_2\,$; therefore a ${\cal Z}_2$ string
is formed during this phase transition.  The subsequent symmetry
breakings can be implemented by the adjoint {\bf 45} of SO(10) and
the fundamental {\bf 10} in the usual fashion:
\begin{eqnarray}
        \hbox{Spin(10)} &\stackrel{\bf 126}{\longrightarrow}&
        \hbox{SU(5)}\times{\cal Z}_2 \nonumber\\
        & \stackrel{\bf 45}{\longrightarrow} &
        \hbox{SU(3)}\times\hbox{SU(2)}\times\hbox{U(1)}
        \times{\cal Z}_2             \nonumber\\
        & \stackrel{\bf 10}{\longrightarrow} &
        \hbox{SU(3)}\times\hbox{U(1)}_{\hbox{em}}\times{\cal Z}_2\,.
\end{eqnarray}
This ${\cal Z}_2$ string survives all the symmetry breakings
since ${\cal Z}_2$ is preserved at low energies.

The {\bf 126} representation consists of fifth\--rank
anti-symmetric tensors satisfy\-ing the self\--duality condition
\begin{equation}
   \phi_{i_1...i_5} = \frac{i}{5!}  \epsilon_{i_1....i_{10}}
                        \phi_{i_6...i_{10}}.
\end{equation}
The component which acquires an expectation value $\langle\phi
\rangle$ transforms as an SU(5) singlet, and to write it down
explicitly, we first specify how the SU(5) subgroup is embedded
in SO(10).  The fundamental representation of SO(10) consists of
10$\times$10 matrices, which can be labeled by indices $i, i = 1,
\ldots ,10\,.$  The generators of SO(10) in this representation
can be written as antisymmetric, purely imaginary matrices. The
generators of SU(5) in the fundamental representation are hermitian,
traceless 5$\times$5 matrices which can be written as
\begin{equation}
     \tau_{\alpha \beta} = S_{\alpha \beta} + iA_{\alpha \beta}\,,
\end{equation}
where $\alpha,\beta =1,..,5$ label the matrix elements, and $S, A$
are real 5$\times$5 matrices, representing the real and imaginary
parts of $\tau$. Hermiticity and tracelessness of $\tau$ require
$S_{\alpha \beta} = S_{\beta \alpha}, A_{\alpha \beta} =
-A_{\beta\alpha}$, and $TrS=0$.  A natural way to embed SU(5)
in SO(10) is to treat five-dimensional complex vectors as
ten-dimensional real vectors, {\it i.e.} replace the paired
indices ($\alpha, a$), where $\alpha = 1, \ldots ,5$ label a
five-dimensional vector and $a=1,2$ label its real and imaginary
parts, by the index $i,\,i=1, \ldots ,10$.  Then, the generators
of the subgroup SU(5) of SO(10) can be expressed as
\begin{equation}
        \tau_{\alpha a,\,\beta  b} = i( A_{\alpha \beta}I_{ab} +
               S_{\alpha \beta}M_{ab})\,,
\end{equation}
where $I$ is the 2$\times$2 identify matrix and $M = i \sigma_2\,,
\sigma_2$ being the second 2$\times$2 Pauli matrix.  One can
convince oneself that in this $(\alpha, a)$ notation, the
rank-five antisymmetric Levi-Civita tensor
$\epsilon_{\alpha_1 \alpha_2 \alpha_3 \alpha_4 \alpha_5 }$ which
transforms as an SU(5) singlet in the SU(5) notation becomes
\begin{equation}
         i^{f(a_1...a_5)} \epsilon_{\alpha_1\alpha_2\alpha_3
               \alpha_4\alpha_5}\,,
\end{equation}
where $f(a_1 \ldots a_5)$ is defined to equal the number of $a_i$
that takes the value 2.  It is also straightforward to check that
this expression satisfies the self-duality condition (Eq.~(3)).
Thus $\langle\phi\rangle$ is written as
\begin{equation}
     \langle \phi_{\alpha_1 a_1...\alpha_5 a_5} \rangle
        = \mu\ i^{f(a_1...a_5)}
        \epsilon_{\alpha_1 \alpha_2 \alpha_3 \alpha_4 \alpha_5 }\,,
\end{equation}
where $\mu$ is a parameter.

Some words about our notation.  The tensor indices $i_1,\ldots ,i_5$
of $\phi_{i_1 \ldots i_5}$ will be suppressed for convenience and
legibility whenever no ambiguity should arise.  In the expressions
like $\tau\phi$ and $e^{i\tau\theta} \phi$ where $\tau$ operates
on $\phi$, $\tau$ is understood to be in the same representation
of $\phi$, {\it i.e.} $\tau$ is the short-hand for
\FL
\begin{equation}
   \tau_{i_1 \ldots i_5j_1 \ldots j_5} =
	\tau_{i_1j_1} \delta_{i_2j_2} \ldots \delta_{i_5j_5}
	+ \delta_{i_1j_1} \tau_{i_2j_2} \ldots \delta_{i_5j_5}
	+  \ldots
\end{equation}

With the symmetry breaking Spin(10)$\rightarrow$SU(5)$
\times{\cal Z}_2$, strings are formed.  At spatial infinity,
the general form of $\phi$ is given by Eq.~(1). For the energy
to be finite, the co\-variant derivative of $\phi$,
$D_\mu \phi \equiv \partial_\mu \phi + eA_\mu \phi\ $, has to
vanish at spatial infinity; therefore the gauge field $A_\mu$
takes the form $A^\theta = i\frac{1}{er} \tau$, $A^r = 0\,,$ as
$r \rightarrow \infty$. In the core of the string, there is a
magnetic flux $\oint \vec{A} \cdot d\vec{l} = \frac{2\pi}{e}\tau$
pointing in the direction of $\tau$ in group space.  Strings
carrying flux pointing in different directions in group space
are topologically equivalent since the only nontrivial winding
number here is one, but dynamically they can differ.  Because
the scalar field $\phi(\theta)$ varies with $\theta$, the
embedding of the unbroken subgroup SU(5) in SO(10) outside
the string also varies with $\theta$.  More precisely,
the generators $\tau^a_\theta, a=1, \ldots ,24$ of the unbroken
SU(5) at $\theta$ are related to the generators $\tau^a_0$ of
the unbroken SU(5) at $\theta=0$ by the similarity transformation
\begin{equation}
        \tau^a_{\theta}= g(\theta)\tau_0^a g^{-1}(\theta)\,,\ \
        g(\theta)=e^{i\tau\theta}\,.
\end{equation}
Consequently, the fermion fields which transform as {\bf 1},
$\bf{\bar 5}$ and {\bf 10} under SU(5) are also rotated as
one goes around the string.  How the fields mix depends on
which direction in group space $\phi(\theta)$ winds.

The SO(10) generators can be written as 10$\times$10 matrices
of the form $(\tau^{ab})_{ij} = -i(\delta^a_i \delta^b_j
- \delta^b_i\delta^a_j)\,,$ where $a,b$ label the group
indices, $i,j$ label the matrix elements, and $a,b,i,j$ all
run from 1 to 10.  In this notation $\tau_{\rm all}$ is given by
\begin{equation}
 \tau_{\rm all}\equiv \frac{1}{5} (\tau^{12} +
	\tau^{34} + \ldots +\tau^{9\,10})\,,
\end{equation}
where the factor of 1/5 is included for $\phi(\theta)$ to have a
$2\pi$ rotational period.  It takes a little more effort to write
down the $\tau_1$'s.  Let us first write the SU(5) generators
specified by Eq.~(5) in terms of $\tau^{ab}$ given above.
The four diagonal generators are trivial.  For the other twenty
generators, one can group the 10$\times$10 space into 2$\times$2
blocks, and write the 45 $\tau^{ab}$'s as $\tau^{2\alpha-1,\,
2\beta-1}, \tau^{2\alpha-1,\,2\beta}, \tau^{2\alpha,\, 2\beta-1}$
and $\tau^{2\alpha, 2\beta}$, where $\alpha, \beta$ both run from
1 to 5.  Then it is not hard to see that the twenty linear
combinations
\begin{eqnarray}
         && \frac{1}{2} (\tau^{2\alpha-1,\,2\beta}
        -\tau^{2\alpha,\,2\beta-1})\,,\nonumber\\
         && \frac{1}{2} (\tau^{2\alpha-1,\,2\beta-1}
        +\tau^{2\alpha,\,2\beta})\,,\quad \alpha < \beta
\end{eqnarray}
are all of the form of Eq.~(5), and therefore can be chosen
to be the twenty off-diagonal generators of SU(5). Note that
the superscripts $\alpha, \beta$ above label the group indices
while the subscripts $\alpha, \beta$ in Eq.~(5) label the
matrix elements.  The twenty $\tau_1$'s outside SU(5) then can
be expressed by the other twenty linear combinations as
\begin{eqnarray}
         \tau_1 &\equiv& \frac{1}{2}(\tau^{2\alpha-1,\,2\beta}
                        +\tau^{2\alpha,\,2\beta-1})\,,\nonumber\\
        && \frac{1}{2}(\tau^{2\alpha-1,\,2\beta-1} -
           \tau^{2\alpha,\,2\beta})\,,\quad \alpha < \beta\,.
\end{eqnarray}

Other than the SU(5) group properties, the linear combinations
above can also be classified under the group SO(4), which is
locally isomorphic to SU(2)$\times$SU(2).  For a given $\alpha$
and $\beta$ where $\alpha < \beta$, the two generators of
Eq.~(11) plus the diagonal
\begin{equation}
  \frac{1}{2} (\tau^{2\alpha-1,\,2\beta-1}-\tau^{2\alpha,\,2\beta})
\end{equation}
can be easily shown to obey the SU(2) algebra.  Similarly,
the two generators of Eq.~(12) plus
\begin{equation}
  \frac{1}{2} (\tau^{2\alpha-1,\,2\beta-1}+\tau^{2\alpha,\,2\beta})
\end{equation}
generate another SU(2).  Thus, for a given $\alpha$ and $\beta$
$(\alpha < \beta)$, the six generators of Eqs.~(11-14) generate
rotations in the 4-dimensional space spanned by vectors in the
$2\alpha-1, 2\alpha, 2\beta-1, 2\beta$ directions.

\section{Field Configurations}
The relevant part of the Lagrangian for the SO(10) theory is
given by
\begin{equation}
         {\cal L} = \frac{1}{4} trF_{\mu \nu} F^{\mu \nu} +
                (D_\mu \phi )^\ast (D^\mu \phi) - V(\phi)
\end{equation}
where $F_{\mu \nu} = -iF_{\mu \nu}^a \tau_a\,, A_{\mu} = -iA_{\mu}^a
\tau_a\,, F_{\mu \nu} = \partial_\mu A_\nu - \partial_\nu A_\mu +
e[A_\mu ,A_\nu ]\,, D_\mu = \partial_\mu + e A_\mu\ $;
$A^a_\mu, a=1, \ldots ,45$, are the SO(10) gauge fields and
$\phi$ is the Higgs {\bf 126}.   The most general gauge-invariant
and renormalizable potential $V(\phi)$ contains all the distinct
contractions of two and four $\phi$'s:
\FL
\begin{eqnarray}
V(\phi) & = & v_1 \phi_{i_1 \ldots i_5} \phi^{\ast}_{i_1 \ldots i_5}
       +   v_2 (\phi_{i_1 \ldots i_5} \phi^{\ast}_{i_1 \ldots i_5})^2
                  \nonumber\\
   & + & v_3 \phi_{i_1 n_2 n_3 n_4 n_5}
             \phi^\ast_{j_1 n_2 n_3 n_4 n_5}
             \phi_{i_1 \ell_2 \ell_3 \ell_4 \ell_5}
	     \phi^\ast_{j_1 \ell_2 \ell_3 \ell_4 \ell_5} \nonumber\\
   & + & v_4 \phi_{i_1 i_2 n_3 n_4 n_5}
             \phi^\ast_{j_1 j_2 n_3 n_4 n_5}
	     \phi_{i_1 i_2 \ell_3 \ell_4 \ell_5}
	     \phi^\ast_{j_1 j_2 \ell_3 \ell_4 \ell_5} \nonumber\\
   & + & v_5 \phi_{i_1 j_2 n_3 n_4 n_5}
             \phi^\ast_{j_1 i_2 n_3 n_4 n_5}
	     \phi_{i_1 i_2 \ell_3 \ell_4 \ell_5}
	     \phi^\ast_{j_1 j_2 \ell_3 \ell_4 \ell_5} \nonumber\\
   & + & v_6 \phi_{i_1 i_2 j_3 n_4 n_5}
                  \phi^\ast_{j_1 j_2 i_3 n_4 n_5}
		  \phi_{i_1 i_2 i_3 \ell_4 \ell_5}
		  \phi^\ast_{j_1 j_2 j_3 \ell_4 \ell_5}\,.\ \
\end{eqnarray}
In writing down the $v_3$ through $v_6$ terms above, one has to
consider two things: (1) the possible ways to contract the
indices, and (2) which $\phi$'s are to be complex conjugated.
One can deal with (1) without the complication of (2) by adopting
an equivalent real 252 representation for $\phi$ because a complex,
self-dual 126-dimensional tensor can be thought of as a real,
252-dimensional tensor by dropping the self-duality condition
and taking the real parts of the resulting complex, 252-dimensional
tensor.  One can see there are only four distinct terms and they
are terms $v_3$ through $v_6$ in Eq.~(16) above.  Then when $\phi$
is taken to be complex, two out of the four $\phi$'s have to be
complex conjugated to make the potential real.  There are three
possibilities: $\phi\phi^{\ast} \phi\phi^{\ast},
\ \phi^{\ast}\phi\phi\phi^{\ast},
\ \phi\phi \phi^{\ast}\phi^{\ast}\ $, for each of the four
contractions $\phi\phi\phi\phi$ when $\phi$ is real.  But
after the self-duality condition is applied, one can show
that only one of the three terms is actually independent.

The Euler-Lagrange equations of motion for $\phi$ and $A_\mu$
are given by
\begin{eqnarray}
   && D_\mu D^\mu \phi = -\frac{\partial V}{\partial \phi^\ast}\,,
        \label{eq:EOMI}\\
   && Tr(\tau^{a\,2})(\partial_\mu F^{a\,\mu \nu} +
                ef^{abc} A_\mu ^b F^{c\,\mu\nu}) \nonumber\\
   && \qquad = ie\{(D^\nu \phi)^\ast (\tau^a \phi) -
	(\tau^a \phi)^\ast (D^\nu \phi)\} \,,
\end{eqnarray}
where $a$ is not summed over, and where a basis has been chosen
so that $Tr(\tau^a \tau^b)=\delta^{ab} Tr(\tau^{a\,2})$.

We construct for string-$\tau_{\rm all}$ a solution of
the following form:
\newline {\em Ansatz I\ }:
\begin{eqnarray}
        \phi & = & f(r) e^{i\tau_{\rm all}\theta} \phi_0
               =   f(r) e^{i\theta} \phi_0\,, \nonumber\\
        A^\theta & = & i\frac{g(r)}{er} \tau_{\rm all}\,,
	\label{eq:ansI}\\
        A^r & = & 0\,, \nonumber \
\end{eqnarray}
where $\phi_0 \equiv \langle\phi\rangle$ as defined in
Eq.~(7). The boundary conditions on the functions are
\begin{eqnarray}
  f(0) = 0\,,\qquad &
  f(r) \stackrel{r\rightarrow \infty}{\longrightarrow} \mu\,,
                \nonumber\\
  g(0) = 0\,,\qquad &
  g(r) \stackrel{r\rightarrow\infty}{\longrightarrow} 1\,;
\end{eqnarray}
$V(\phi)$ is minimized at $f=\mu$.  Inserting this {\it ansatz\ }
into the equations of motion and using the relations $\tau_{\rm all}
\tau_{\rm all} \phi_0 = \phi_0\ $ and $(\tau_{\rm all}\phi_0)^\ast
(\tau_{\rm all}\phi_0) = \phi_0^\ast \phi_0 = 3840 \equiv N$,
we obtain two coupled differential equations for $f(r)$ and $g(r)$:
\begin{eqnarray}
        f^{\prime\prime} + \frac{1}{r} f^\prime
          - \frac{(1-g)^2}{r^2} f & = & f(v_1+2Nv_2 f^2)
                \,,\nonumber\\
        Tr(\tau_{\rm all}^2) \left( g^{\prime\prime} - \frac{1}{r}
                g^\prime \right)  & = & -2N e^2 (1-g) f^2 \,,
\end{eqnarray}
where the prime denotes differentiation with respect to $r$, and
$Tr(\tau_{\rm all}^2) = \frac{2}{5}$ from Eq.~(10).  An expansion
of $f(r)$ and $g(r)$ in powers of $r$ around the origin reveals
that $f(r)$ is odd in $r$ with a linear leading term, whereas
$g(r)$ is even in $r$ with a quadratic leading term.

Inserting {\it Ansatz I\ } for string-$\tau_{\rm all}$ into the
Lagrangian gives
\begin{eqnarray}
   -{\cal L}^{\rm all} &=& \frac{Tr(\tau_{\rm all}^2)}{2e^2 r^2}
	g^{\prime\,2}
    + N f^{\prime\,2} + N \frac{(1-g)^2}{r^2} f^2 \nonumber\\
   &&  + N(v_1 f^2 + Nv_2 f^4)\,.
\end{eqnarray}
As a consistency check, note that the equations of motion
obtained by varying ${\cal L}^{\rm all}$ with respect to the
functions $g$ and $f$ are identical to those in Eq.~(21).

Note that the parameters $v_3$ through $v_6$ in the potential
$V$ are absent from Eq.~(21) and ${\cal L}^{\rm all}$ above.
This is because whenever one index of a given $\phi$ is
contracted with one index of another $\phi$, this index is
summed over from 1 through 10, or in the $(\alpha, a)$ notation
discussed earlier, from $\alpha = 1$ through 5 and $a=1,2$.  For
a given $\alpha$, the term with $a=2$ by definition has an extra
factor of $i^2=-1$ compared to the term with $a=1$. These two terms
cancel each other when they are added.  Because this is true for
every $\alpha$, the third through the sixth terms in $V$ vanish
identically for the string-$\tau_{\rm all}$ {\it ansatz}.

To construct an {\it ansatz\ } for string-$\tau_1$, we need to
consider separately the two sets of generators in Eq.~(12),
which will be referred to as
\begin{eqnarray}
         \tau_{1+} &=& \frac{1}{2}(\tau^{2\alpha-1,\,2\beta}
                        +\tau^{2\alpha,\,2\beta-1})\,, \nonumber\\
         \tau_{1-} &=& \frac{1}{2}(\tau^{2\alpha-1,\,2\beta-1}
                        -\tau^{2\alpha,\,2\beta})\,,
        \ \ \alpha < \beta\,.
\end{eqnarray}
As we shall see, it is sufficient to derive the equations of motion
for an {\it ansatz\ } based on a generator of the form $\tau_{1+}$.
By a simple redefinition, it will then be possible to construct
an {\it ansatz\ } based on a generator of the form $\tau_{1-}$.
For now, we consider the case when $\tau_1$ has the form $\tau_{1+}$.
The simple extension of {\it Ansatz I} with $\tau_{\rm all}$
replaced by $\tau_1$ does not work for string-$\tau_1$.
The problem arises from the term $\tau_1\tau_1 \phi$ on the
left-hand side of Eq.~(17) in which a new tensor $\phi_0^A$,
\begin{equation}
        \tau_1\tau_1 \phi_0 =  \phi^A_0 \,,
\end{equation}
is generated, where
\FL
\begin{equation}
        \phi^A_{0\,i_1 \ldots i_5} \equiv
        \left\{ \begin{array}{ll}
             \phi_{0\,i_1 \ldots i_5}\,,\ &
                \mbox{if two indices take the values} \\
              & \mbox{$(2\alpha-1, 2\beta-1)$
                   or $(2\alpha, 2\beta)$}\,,\\
              0     \,, &  \mbox{otherwise}.
        \end{array} \right.
\end{equation}
As a result, the differential equations for $g(r)$ and $f(r)$
are satisfied only if $g(r)=1$ or $f(r)=0$ everywhere, which
is not consistent with the boundary conditions given by Eq.~(20).
(Note that the solution $g=1$ and $f=\mu$ is the vacuum field
configuration expressed in a singular gauge.)

We construct a nontrivial solution for string-$\tau_1$ by
replacing $f(r)\phi_0$ and $\tau_{\rm all}$ in {\it Ansatz I}
with $(f_1(r)\phi_0 + f_2(r)\phi^A_0)$ and $\tau_1$ respectively.
Note that $\phi_0$ is not orthogonal to $\phi^A_0$ because
$\phi^A_{0\,i_1 \ldots i_5} \phi^\ast_{0\,i_1 \ldots i_5} \neq 0$.
Therefore instead of expanding $\phi$ in $\phi_0$ and $\phi^A_0$,
we will use the more convenient basis $\phi^A_0$ and $\phi^B_0$
where
\begin{equation}
        \phi^B_0 \equiv \phi_0 - \phi^A_0\
\end{equation}
and $\phi^B_0$ is orthogonal to $\phi^A_0$:
\begin{equation}
   \phi^A_{0\,i_1 \ldots i_5} \phi^{B\,\ast}_{0\,i_1 \ldots i_5}
	= 0\,.
\end{equation}
{}From the definition of $\phi^A_0$ (Eq.~(25)) and the properties of
$\phi_0$, one can see that
\FL
\begin{equation}
        \phi^B_{0\, i_1 \ldots i_5} =
        \left\{ \begin{array}{ll}
            \phi_{0\,i_1 \ldots i_5}\,,\ &
    \mbox{if two indices take the values} \\
   & \mbox{$(2\alpha-1, 2\beta)$ or $(2\alpha, 2\beta-1)$} \,,\\
             0     \,, &  \mbox{otherwise}
        \end{array} \right.
\end{equation}
and $\phi^B_0$ is annihilated by $\tau_1$:
\begin{equation}
        \tau_1  \phi^B_0 = 0\,.
\end{equation}
The solution constructed for string-$\tau_1$ is
\newline {\em Ansatz II\ }:
\begin{eqnarray}
        \phi & = & e^{i\tau_1 \theta} \left\{ f_o(r) \phi^A_0 +
                    f_e(r) \phi^B_0 \right\} \,, \nonumber\\
        A^\theta & = & i\frac{g(r)}{er} \tau_1 \,, \\
        A^r & = & 0\,, \nonumber
\end{eqnarray}
where as will become clear in the next two paragraphs, the
functions $f_o(r)$ and $f_e(r)$ are named after their odd
and even parities in $r$.

At the origin, we require the fields to be regular.  Since
$\phi^B_0$ is left invariant by $e^{i\tau_1\theta}$ (Eq.~(29))
but $\phi^A_0$ is not, at the origin $f_e(0)$ can be any
constant but $f_o(0)$ has to vanish.  At large $r$,
the scalar field $\phi$ has to take the form
\begin{equation}
        \phi \stackrel{r\rightarrow \infty}{\longrightarrow} \mu
        \ e^{i\tau_1 \theta} \phi_0 = \mu\ e^{i\tau_1 \theta}
        (\phi^A_0 + \phi^B_0)
\end{equation}
for the unbroken gauge group to be SU(5), so both $f_o(r)$
and $f_e(r)$ approach $\mu$ at large $r$.  The boundary
conditions on the functions are
\begin{eqnarray}
     &  f_o(0)=0\,,\qquad &
     f_o(r) \stackrel{r\rightarrow \infty}{\longrightarrow} \mu\,,
        \nonumber\\
     &  f_e(0)= a_0\,,\qquad &
     f_e(r) \stackrel{r\rightarrow \infty}{\longrightarrow} \mu\,,
        \nonumber\\
     &  g(0)=0\,,\qquad &
     g(r)\stackrel{r\rightarrow \infty}{\longrightarrow} 1\,,
\end{eqnarray}
where $a_0$ is a constant.

The equations of motion for $\phi$ and $A_\mu$ are closed when
the fields take the form in {\it Ansatz II\ }.  We obtain three
coupled differential equations for $f_o(r),f_e(r)$ and $g(r)$.
The algebra involved in extracting these three equations, however,
is considerably more tedious than in the $\tau_{\rm all}$ case
mainly because the forms of $\phi^A_0, \phi^B_0$ and $\tau_1$ are
less symmetric. We will not present the algebra involved and simply
quote the results:
\FL
\begin{eqnarray}
        f_e^{\prime\prime} + \frac{1}{r} f_e^\prime
        & = & f_e \left\{ v_1 + N v_2 (f_o^2 + f_e^2)
         -\frac{N}{25} e^2 \lambda_3 (f_o^2 - f_e^2) \right\}
		\nonumber\\
        f_o^{\prime\prime} + \frac{1}{r} f_o^\prime
        & - & \frac{(1-g)^2}{r^2} f_o \nonumber\\
        & = & f_o \left\{ v_1 + N v_2 (f_o^2 + f_e^2)
          + \frac{N}{25} e^2 \lambda_3 (f_o^2 - f_e^2) \right\}
	\nonumber
\end{eqnarray}
\FL
\begin{equation}
        Tr(\tau_1^2) \left( g^{\prime\prime}  -
        \frac{1}{r} g^\prime \right) = -N e^2 (1-g) f_o^2 \,,
\end{equation}
where $e^2 \lambda_3 \equiv  v_3 + \frac{v_4}{4} + \frac{v_5}{4}
+ \frac{v_6}{12}$, and $Tr(\tau_1^2)=1$ from Eq.~(12).  An expansion
of $g, f_o$ and $f_e$ in powers of $r$ around the origin gives
\begin{eqnarray}
  f_o(r) & = & a_1 r + a_3 r^3 + \ldots \,,\nonumber\\
  f_e(r) & = & a_0 + a_2 r^2 + \ldots \,,\nonumber\\
  g(r)   & = & b_2 r^2 + b_4 r^4 + \ldots \,,
\end{eqnarray}
where the coefficients of all the higher terms are related to
$a_0, a_1$ and $b_2$ recursively.  The function $f_o$ is indeed
odd and $f_e$ even in $r$ as claimed earlier.

Inserting {\it Ansatz II\ } for string-$\tau_1$ into the
Lagrangian gives
\FL
\begin{equation}
 -{\cal L}^1 = \frac{Tr(\tau_1^2)}{2e^2 r^2} g^{\prime\,2}
    + \frac{N}{2} \left( f_e^{\prime\,2} + f_o^{\prime\,2} \right)
        + \frac{N}{2} \frac{(1-g)^2}{r^2} f_o^2 + V_{ans}
\end{equation}
where
\begin{eqnarray}
        V_{ans} &=& \frac{N}{2} \left\{ v_1 (f_o^2 + f_e^2)
         + \frac{N}{2} v_2 (f_o^2 + f_e^2)^2 \right. \nonumber\\
  && \left. +\frac{N}{50} e^2 \lambda_3 (f_o^2 - f_e^2)^2 \right\}\,.
\end{eqnarray}
Here again, note that the equations of motion obtained
by varying ${\cal L}^1$ with respect to the functions
$g, f_o$ and $f_e$ are identical to those in Eq.~(33).

Now let us consider the other case when $\tau_1$ has the form of
$\tau_{1-}$.  One can show that Eq.~(24) now is $\tau_1\tau_1\phi_0
=\phi^B_0$, and instead of $\tau_1 \phi^B_0=0$, one has
$\tau_1 \phi^A_0=0$.  Therefore by switching the definitions of
$\phi^A_0$ and $\phi^B_0$ in Eqs.~(25) and (28), all the equations
between (24) and (32) are preserved, and one can show that the
equations of motion are unchanged.  We conclude that {\it Ansatz II}
applies to all twenty $\tau_1$'s, where for $\tau_{1+}$, $\phi^A_0$
and $\phi^B_0$ are defined by Eqs.~(25) and (28) respectively, but
for $\tau_{1-}$, the definitions of the two are reversed.
The equations of motion are given by Eq.~(33) for all cases.

\section{Numerical Calculations}
In this section we present the numerical solutions to the two
sets of differential equations (21) and (33) with the appropriate
boundary conditions at the origin and some large value of $r$.
We implemented two methods: the ``shooting'' and the relaxation
methods to handle this two-point boundary value problem.  In the
``shooting'' method \cite{num rec}, an initial guess for the free
parameters at $r=0$ was made and then the equations were integrated
out to large $r$ where the boundary conditions were specified.  As
the name of the method suggests, the true solutions were found by
adjusting the parameters at $r=0$ in the beginning of each iteration
to reduce the discrepancies from the desired boundary conditions at
large $r$ computed in the previous iteration.  For string-$\tau_1$,
the small-$r$ expansion of the functions in Eq.~(34) gives
$g(0) = g^\prime(0) = 0\,, f_o(0) = f_o^{\prime\prime}(0) =
f_e^\prime(0) = 0\,$, and $f_e^{\prime\prime}(0)=2a_2\,,$ where
$a_2$ is related to $a_0$, $a_1$ and $b_2$, but the values of
\begin{eqnarray}
     f_e(0) &=& a_0\,, \nonumber\\
     f_o^\prime(0) &=& a_1\,, \nonumber\\
     g^{\prime\prime}(0) &=& 2b_2\,,
\end{eqnarray}
were adjusted to match the boundary conditions at large $r$.
For string-$\tau_{\rm all}$, we have shown that $f(r)$ is odd and
$g(r)$ is even in $r$, with $f(r)=ar+\ldots$ and $g(r)=br^2+\ldots$.
Thus only the two values $f^\prime(0), g^{\prime\prime}(0)$ were free
parameters.  At large $r$, discrepancies from the boundary condition
were corrected by the multi-dimensional Newton-Raphson method which
computed the corrections to the initial parameters.  With an initial
guess for the parameters at $r=0$, this ``shooting'' process was
iterated until the ``targets'' were met.  The fourth-order
Runge-Kutta method was used to integrate the equations.

We have also implemented a relaxation scheme for comparison.
In this method the first step is to express the string
energy as a function of the values of the functions $f$ and
$g$ (or $f_e$, $f_o$, and $g$) defined on an evenly spaced
mesh of points.  While a Simpson's rule approximation worked
well for the middle range of parameters, a more sophisticated
approximation was used to extend the range of parameters that
could be treated.  For each interval of two lattice spacings,
smooth functions $\tilde f$ and $\tilde g$ were defined by 2nd
order polynomial interpolation from the three mesh points
(midpoint and two end points); with the help of a symbolic
integration program, the integral defining the energy was
carried out exactly for the interpolated functions.  (By this
method the energy obtained is a rigorous upper limit on
the true ground state string energy.)  To avoid divergences
caused by the explicit factors of $1/r^2$ in the energy density,
the first interval had to be treated more carefully--- instead
of fitting the functions with a 2nd order polynomial, we fitted
the coefficients of the analytically determined power series,
such as Eq.~(34).  Trial functions $f$ and $g$ were chosen,
and then the energy was minimized by varying each mesh point
one at a time, successively going through the lattice many
times.  We found it efficient to begin with a coarse mesh
which was made successively finer by factors of 2,
interpolating the solution at each stage to obtain the first
trial solution for the next stage.  For the final run in
each case we used 2048 points.

We found the results by the two methods to agree to
approximately one part in a million or better.  In general we
were able to explore a wider parameter range with the relaxation
method than with the ``shooting'' method, but the qualitative
features given by the ``shooting'' method remained the same.
(The author wishes to thank Alan Guth for implementing the
relaxation part of the calculations.)

The dependence of the equations on the parameters in the theory can
be simplified if $r, f, f_o$ and $f_e$ are rescaled as ($v_1 < 0$)
\begin{eqnarray}
     r & \rightarrow & \sqrt{-v_1} r\,, \nonumber\\
     \{f\,, f_o\,, f_e\} & \rightarrow & \sqrt{\frac{2Nv_2}{-v_1}}
                        \{f\,, f_o\,, f_e\}\,.
\end{eqnarray}
Then only the following combinations of parameters appear
in the differential equations:
\begin{eqnarray}
    \lambda_2 & \equiv & \frac{v_2}{e^2} \,,\nonumber\\
    \lambda_3 & \equiv & \frac{1}{e^2} \left(
     v_3 + \frac{v_4}{4} + \frac{v_5}{4} + \frac{v_6}{12}\right)\,.
\end{eqnarray}
The Hamiltonian densities ${\cal H}^{\rm all}$ and ${\cal H}^1$ for
the two strings are simply $-{\cal L}^{\rm all}$ and $-{\cal L}^1$
given by Eqs.~(22) and (35) because all fields are assumed to be
time-independent.  With the same rescaling, one obtains
\begin{eqnarray}
  \frac{v_2}{(-v_1)^2} {\cal H}^{\rm all} &=& \frac{1}{2} \left\{
   \frac{2\lambda_2}{5r^2} g^{\prime\,2}
       + f^{\prime\,2} + \frac{(1-g)^2}{r^2} f^2 \right.\nonumber\\
  && \left. + \frac{1}{2}(1-f^2)^2 \right\}
\end{eqnarray}
and
\begin{eqnarray}
        &&\frac{v_2}{(-v_1)^2} {\cal H}^1 = \frac{1}{2} \left\{
        \frac{\lambda_2}{r^2} g^{\prime\,2}
        + \frac{ f_o^{\prime\,2} + f_e^{\prime\,2}}{2}
        + \frac{(1-g)^2}{2r^2} f_o^2  \right.   \nonumber\\
        &&\ + \left.
         \frac{1}{2} \left( 1 - \frac{f_o^2 + f_e^2}{2}  \right)^2
        + \frac{\lambda_3}{200\lambda_2} (f_o^2 - f_e^2)^2 \right\}
\end{eqnarray}
where the $\tau_{\rm all}$ equation depends on $\lambda_2$ only but
the $\tau_1$ equation depends on both $\lambda_2$ and $\lambda_3$.

Typical solutions for the two strings calculated from the
``shooting'' method are shown in Figs.~1 and 2, where
$\lambda_2 = 0.132$ and $\lambda_3 = 10.25$.  For the same
$\lambda_2$ and $\lambda_3$, the solutions given by the
relaxation method appear indistinguishable visually from
those in Figs.~1 and 2.  For string-$\tau_{\rm all}$, we
were able to find solutions in the approximate range $10^{-2}
< \lambda_2 < 10$ using the ``shooting'' method and $10^{-4}
< \lambda_2 <10^3$ using the relaxation method.  For
string-$\tau_1$, we explored the range $5\times 10^{-2} <
\lambda_2 < 1$ and $0.5 < \lambda_3 < 10^2$.  In general,
the functions converged more slowly near the two ends of each
range above, and we did not attempt to find solutions beyond
these limits.  We numerically integrated ${\cal H}^{\rm all}$
and ${\cal H}^1$ for the solutions we computed, and found
string-$\tau_1$ to have the lower energy for all the parameters
we explored.  In Fig.~3, the energy density $2\pi r {\cal H}$
of the two solutions shown in Figs.~1 and 2 is plotted, and
the energy of string-$\tau_1$ is clearly lower.  For comparison,
we point out that the energy per unit length of
string-$\tau_{\rm all}$ in the range $0.9 < \lambda_2 < 4.0$ has
been calculated by Aryal and Everett \cite{everett}.  Our values
in this range of parameters agree with theirs to within 1\%.

One of the most important properties of the two strings we
investigate in this paper is whether string-$\tau_1$ has
lower energy than string-$\tau_{\rm all}$.  We just showed
that this is true for some range of the parameters.  To
systematically explore a wider parameter range, however, it
is very laborious and time-consuming to calculate the $\tau_1$
solutions for different $\lambda_2$ and $\lambda_3$ first and
then compute the corresponding energy.  Instead, we employ an
upper-bound argument to reduce the two-dimensional parameter space
$(\lambda_2, \lambda_3)$ to one.  We set $f_o = f_e \equiv f_1$ in
the Lagrangian and take $g(r), f_1(r)$ as trial functions for
string-$\tau_1$.  The advantage in using $f_o = f_e$ is that the
last term in Eq.~(41) vanishes, and the equations no longer depend
on $\lambda_3$.  Moreover, Eqs.~(40) and (41) then have the same
functional form, differing only in the coefficients of the first
and the third terms, and one can solve the equations for
string-$\tau_1$ the same way as for string-$\tau_{\rm all}$ using
different values of $\lambda_2$.  The corresponding energy,
denoted by $E_1(f_o=f_e)$, gives an upper bound on the true
energy of string-$\tau_1$ by the variational principle.
If $E_1(f_o=f_e) < E_{\rm all}$ for a given $\lambda_2$, then one
can conclude that string-$\tau_1$ has the lower energy for that
value of $\lambda_2$ and all values of $\lambda_3$.
(Note that in the limit of $\lambda_3 \rightarrow \infty$, the
trial functions approach the true string solution because for the
energy to be finite, the last term in Eq.~(41) requires $f_o
\rightarrow f_e$.)  Our result is presented in Fig.~4, where the
ratio $E_1(f_o=f_e)/E_{\rm all}$ is plotted as a function of
log$\,\lambda_2$ for $10^{-4} < \lambda_2 < 2.5\times 10^3$.
Note that $E_1(f_o=f_e)/E_{\rm all} < 1$ for all 7 decades of
$\lambda_2$, and is approaching an asymptote of 1 (or possibly
less than 1) as $\lambda_2 \rightarrow 0$.  For large $\lambda_2$,
we find the individual curves of $E_{\rm all}$ vs. log$\,\lambda_2$
and $E_1$ vs. log$\,\lambda_2$ approach straight lines,
suggesting that the ratio $E_1(f_o=f_e)/E_{\rm all}$ levels off at a
constant for large $\lambda_2$.  We conclude that string-$\tau_1$ has
lower energy than string-$\tau_{\rm all}$ for $10^{-4} < \lambda_2
< 2.5\times 10^3$ and all $\lambda_3$, and probably is the ground
state for the entire range of the parameters in the theory.

\section{Scattering Solutions}
To study the scattering of fermions by an SO(10) cosmic string,
one first needs to understand the 16-dimensional spinor
representation of SO(10) to which the left-handed fermions
are assigned.  Spinor representations certainly have been discussed
in the literature \cite{spin}, but to establish a common notation,
we discuss in the Appendix the construction of the generators,
the sixteen states and the identification of states with fermions
that are relevant to this paper.

Now we proceed to study the Dirac equation
\begin{equation}
  (i\not\!\partial - e\not\! A^a \tau^a - m)\psi = 0
\end{equation}
in the background fields of string-$\tau_{\rm all}$ and $\tau_1$:
$A^a_\mu \tau^a = A^{\rm all}_\mu \tau_{\rm all}$ and $A^1_\mu
\tau_1$.  As shown in the Appendix, the fermion fields can be
written as a 16-dimensional column vector where each component
is identified with a fermion given by Eq.~(A.16).  The generators
$\tau_{\rm all}$ and $\tau_1$ can be written as 16$\times$16
hermitian matrices, where $\tau_{\rm all}$ is diagonal with one
diagonal entry equal to $\frac{1}{2}$, ten entries equal to
$\frac{1}{10}$ and five entries equal to $-\frac{3}{10}$.  For
$\tau_1$, we choose $\tau_1 = \frac{1}{2} (\tau^{58}+\tau^{67})$ for
illustration.  We find that $\tau_1$ takes the block-diagonal form
\begin{equation}
        -\tau_1 = \frac{1}{2} \left( \begin{array}{cc}
                                B\  & 0\  \\
                                0\  & B\
                        \end{array} \right) \,,
\end{equation}
where
\begin{equation}
        B = \left( \begin{array}{cccc}
                                0 \ & 0 \ & 0 \ & I \\
                                0 \ & 0 \ & 0 \ & 0 \\
                                0 \ & 0 \ & 0 \ & 0 \\
                                I \ & 0 \ & 0 \ & 0
                        \end{array} \right) \,,
\end{equation}
and $I$ is the 2$\times$2 identity matrix.
For string-$\tau_{\rm all}$, since $\tau_{\rm all}$ is diagonal,
Eq.~(42) decouples into sixteen equations, one for each component
of the wave function, and there is no mixing of leptons and quarks
due to twisting of the Higgs.  However, since the sixteen
eigenvalues of $\tau_{\rm all}$ are all fractional, all sixteen
fermions scatter nontrivially off the string via the Aharonov-Bohm
effect.  As pointed out by previous studies, the wave functions of
these fermions can be strongly enhanced near the core of the string,
leading to strong B-violating processes inside the string.

In the case of string-$\tau_1$, upon diagonalizing $\tau_1$ by a
unitary matrix $U$ and simultaneously rotating the fermion basis
$\psi$ in Eq.~(A.16) to $\tilde{\psi} \equiv U\psi$, we can
write $\tilde{\psi}$ as
\FL
\begin{eqnarray}
  \tilde{\psi}
   & = &( e^- + u_1^c\,, e^- - u_1^c\,, \nu^c+d_1\,, \nu^c-d_1\,,
          u^c_2\,, u^c_3\,, d_3\,, d_2\,,   \nonumber\\
     & & u_3 + d_2^c\,, u_3 - d_2^c\,, u_2+d_3^c\,, u_2-d_3^c\,,
             u_1\,, \nu\,, e^+\,, d_1^c)_L \nonumber
\end{eqnarray}
\begin{equation}
	\quad	\quad
\end{equation}
and Eq.~(42) again decouples into sixteen equations of
the form
\begin{equation}
	  (i\not\!\partial + e\lambda_i\not\! A^1 - m)
		\tilde{\psi}_i = 0\,,
\label{eq:dede}
\end{equation}
where each $\tilde{\psi}_i$ interacts with the gauge field with
coupling strength $e\lambda_i\,; \lambda_i$ are the eigenvalues
of $-\tau_1$.  The eigenvalues are $\lambda_i = \frac{1}{2}$ for
$e + u^c_1\,,\nu^c + d_1\,,u_3 + d^c_2\,,u_2 + d^c_3\,, \lambda_i =
 -\frac{1}{2}$ for $e - u^c_1\,,\nu^c -d_1\,,u_3 - d^c_2\,,u_2 -
d^c_3\,,$ and $\lambda_i = 0$ for all others.  Since the $e + u^c$
and $e - u^c$ components have opposite eigenvalues, we expect a pure
$e$ or $u^c$ to turn into a mixture of $e$ and $u^c$ as it
propagates around the string, producing baryon-number violation.

Before calculating the scattering amplitude, we first comment on
the choice of gauge in this problem.  The fields in {\it Ansatz II}
(See Eq.~(30)) for string-$\tau_1$ were constructed in a gauge where
the scalar field $\phi$ winds with $\theta$ and the gauge field
falls off as $r^{-1}$ at large $r$.  The particle content, however,
is probably most transparent in a different gauge where $\phi$ does
not wind with $\theta$ and $A_\mu \rightarrow 0$ at large $r$
everywhere except on a sheet of singularities at $\theta=0$.
We will refer to the former as the $1/r$-gauge and the latter
as the ``sheet'' gauge, in analogy with the ``string'' gauge of
a magnetic monopole.  Continuing to work in the diagonalized basis,
the fermion fields in the ``sheet'' gauge, $\tilde{\psi}_0$,
are related to those in the $1/r$-gauge, $\tilde{\psi}$,
by the gauge transformation
\begin{equation}
    \tilde{\psi}_0 = e^{-i\tau_1 (\pi-\theta)} \tilde{\psi}\,.
\label{eq:gauge}
\end{equation}
We will solve the Dirac equation and calculate the scattering
amplitude in the $1/r$-gauge, and then write down the baryon-number
violating cross section in the ``sheet'' gauge.

In the presence of an infinitely-thin $\tau_1$-string along the
$z$-axis, the gauge field $A^1_\mu$ takes the form
$A^{1\,r} = A^{1\,z}=0, A^{1\,\theta}= \frac{1}{er}\,,$
where $(r, \theta)$ denote the usual polar coordinates with
$\theta$ running counter-clockwise from the positive $x$-axis.
Owing to the symmetry along the $z$-axis, the matrix $\gamma_3$ in
Eq.~(46) drops out, and with the choice for the $\gamma$-matrices
\begin{eqnarray}
        \gamma_0 &= \left( \begin{array}{cc}
                                \sigma_3 &  0  \\
                                   0     &  -\sigma_3
                           \end{array}  \right) \,,
        &\quad \gamma_1 = \left( \begin{array}{cc}
                                i\sigma_2 &  0  \\
                                   0      &  -i\sigma_2
                           \end{array}  \right) \,, \nonumber\\
        \gamma_2 &= \left( \begin{array}{cc}
                                -i\sigma_1 &  0  \\
                                   0       &  i\sigma_1
                           \end{array}  \right) \,,
        &\quad \gamma_3 = \left( \begin{array}{cc}
                                   0\ \ &\ 1  \\
                                  -1\ \ &\ 0
                           \end{array}  \right) \,,
\end{eqnarray}
Eq.~(46) decouples into two independent equations
for the upper and lower 2-component spinors of $\tilde{\psi}_i$,
where the two equations differ by the sign of the mass term.
Writing the upper spinor of $\tilde{\psi}_i$ as
\begin{equation}
        \left( \begin{array}{c}
                \chi_1 (r) \\
                \chi_2 (r) e^{i\theta}
                \end{array}  \right)
                e^{in\theta - iEt} \,,
\end{equation}
one can show
\begin{equation}
    \left( \begin{array}{cc}
     m-E & -i\left( \partial_r + \frac{n+\lambda_i+1}{r} \right) \\
        -i\left( \partial_r - \frac{n+\lambda_i}{r} \right) & -m-E
        \end{array}  \right)
        \left( \begin{array}{c}
                \chi_1  \\
                \chi_2
                \end{array}    \right)  = 0 \,,
\end{equation}
and the solutions are Bessel functions of order $(n+\lambda_i)$
and $-(n+\lambda_i)$:
\FL
\begin{equation}
        \left( \begin{array}{c}
                \chi_1 \\
                \chi_2
                \end{array}  \right) =
        \left( \begin{array}{c}
                J_{\pm(n+\lambda_i)} (kr) \\
                \pm \frac{ik}{E+m} J_{\pm(n+\lambda_i+1)} (kr)
                \end{array}   \right) \,,\ k=\sqrt{E^2-m^2}\,.
\end{equation}
The appropriate boundary conditions to impose, as pointed out
in Ref.~14, are the square-integrability of the wave functions
near the origin and a self-adjoint Hamiltonian.  The usual
requirement that wave functions be regular at the origin is
sometimes too strong and has to be relaxed.  Since $J_\nu (r)
\sim r^\nu / (2^\nu \nu!)$ for small $r$, one can see that
the solutions above are square-integrable if the $+$ sign
is chosen for the modes $n+\lambda_i > 0$, and the $-$ sign for
$n+\lambda_i < -1$.  For the mode $ -1 < n+\lambda_i < 0$,
however, both choices are square-integrable albeit neither is
regular at the origin, and the solution takes the form
\FL
\begin{equation}
  \left( \begin{array}{c}
        \chi_1 \\
        \chi_2
        \end{array}  \right) =
  \left( \begin{array}{c}
      \sin\mu\,J_{n+\lambda_i} + \cos\mu\,J_{-(n+\lambda_i)} \\
      \frac{ik}{E+m}
      (\sin\mu\,J_{n+\lambda_i+1} - \cos\mu\,J_{-(n+\lambda_i+1)})\\
        \end{array}   \right) \,,
\end{equation}
where $\mu$ is the self-adjoint parameter.

The scattering amplitude $f^{\lambda_i}(\theta)$ for the $i$th
fermion in $\tilde{\psi}$ appears in the asymptotic wave function
written as the sum of the incident plane wave and the scattered
part:
\begin{eqnarray}
    \tilde{\psi}_i &\sim &
        u_E e^{-i\lambda_i(\pi-\theta)} e^{i(kx - Et)} \nonumber\\
     && + \sqrt{\frac{i}{r}} v_E e^{-i\lambda_i(\pi-\theta)}
        f^{\lambda_i}(\theta) e^{i(kr - Et)} \,,
\end{eqnarray}
where $u_E$ and $v_E$ are given by
\begin{equation}
    u_E = \left( \begin{array}{c}
                        1  \\
                    \frac{k}{E+m}
                 \end{array} \right) \,, \quad
    v_E = \left( \begin{array}{c}
                        1  \\
                 \frac{k}{E+m} e^{i\theta}
                 \end{array} \right) \,.
\end{equation}
Expanding $e^{ikx}=e^{ikr\cos\theta}$ and $e^{ikr}$ in Bessel
functions using
\begin{equation}
	e^{ikr\cos\theta}
	= \sum_{n=-\infty}^{\infty} i^n J_n(kr) e^{in\theta}\,,
\end{equation}
and with
\begin{equation}
   f^{\lambda_i}(\theta) = \sum_{n=-\infty}^{\infty}
	f_n^{\lambda_i} e^{in\theta}\,,
\end{equation}
Eq.~(53) can be matched to the solutions in Eq.~(51) mode by mode
at large $r$.  Then the scattering amplitude can be calculated:
\begin{equation}
        f^{\lambda_i} (\theta)
        = \frac{i}{\sqrt{2\pi k}} e^{-i([\lambda_i]+1)\theta}
        \left( \frac{ \sin\left( \frac{\theta}{2} - \pi\lambda_i
	\right)} { \sin \frac{\theta}{2} }
         - e^{2i\delta} \right)\,,
\end{equation}
where $[\lambda_i]$ denotes the largest integer less than
or equal to $\lambda_i$, and $\delta$ is related to $\lambda_i$
and the self-adjoint parameter $\tan\mu$ by \cite{gerbert}
\begin{equation}
    \tan \delta
        = \frac{1-\tan\mu}{1+\tan\mu}\,\tan\frac{\lambda_i \pi}{2}\,.
\end{equation}
With the gauge transformation Eq.~(47), one can easily see that
$(\tilde{\psi}_0)_i$ in the ``sheet'' gauge is given by Eq.~(53)
without the phase $e^{-i\lambda_i(\pi-\theta)}$.

To illustrate the processes that violate the baryon number, we
consider an incident beam of electrons propagating in the fields
of the string.  We will study the $(e, u^c)$-subspace and ignore
other fermions since $e$ in $\psi$ is mixed with $u^c$ only.  In
the ``sheet'' gauge, the eigenstates of $\tau_1$ can be written as
\begin{equation}
        e + u^c = \left( \begin{array}{c}
                         1 \\
                         0
                \end{array}  \right) \,,\quad
        e - u^c = \left( \begin{array}{c}
                         0 \\
                         1
                \end{array}  \right) \,,
\end{equation}
and the electron is simply given by
\begin{equation}
        e = \left( \begin{array}{c}
                         \frac{1}{2} \\
                         \frac{1}{2}
                \end{array}  \right) \,.
\end{equation}
An incident wave of electrons can be written as
\begin{equation}
   \tilde{\psi}^e_{0\,inc} = u_E \left( \begin{array}{c}
                                  \frac{1}{2}  \\
                                  \frac{1}{2}
                           \end{array} \right)  e^{i(kx-Et)} \,,
\end{equation}
which scatters into
\FL
\begin{equation}
        \tilde{\psi}_{0\,sca} = \sqrt{\frac{i}{r}} v_E
         \left\{  f^{\frac{1}{2}}(\theta)
                        \left( \begin{array}{c}
                                \frac{1}{2} \\
                                  0
                        \end{array}   \right)
             +    f^{-\frac{1}{2}}(\theta)
                        \left( \begin{array}{c}
                                  0  \\
                                 \frac{1}{2}
                  \end{array} \right) \right\} e^{i(kr-Et)}\,.
\end{equation}
Note that the suppressed index on the 2-component spinors
$u_E$ and $v_E$ should not be confused with the index associated
with the 2-component eigenvectors used here to label the
$e + u^c$ and $e-u^c$ components of the Dirac field.
Rewriting $\tilde{\psi}_{0\,sca}$ above as
\FL
\begin{eqnarray}
        \tilde{\psi}_{0\,sca} &=& \sqrt{\frac{i}{r}} v_E
         \left\{ \left(
        \frac{f^{\frac{1}{2}}(\theta) + f^{-\frac{1}{2}}(\theta)}{2}
                 \right)
                        \left( \begin{array}{c}
                                \frac{1}{2} \\
                                \frac{1}{2}
                        \end{array}   \right)
        \right. \nonumber\\
         && \left.  + \left(
        \frac{f^{\frac{1}{2}}(\theta) - f^{-\frac{1}{2}}(\theta)}
              {2} \right)
                        \left( \begin{array}{c}
                                 \frac{1}{2} \\
                                -\frac{1}{2}
                        \end{array} \right) \right\} e^{i(kr-Et)}\,,
\end{eqnarray}
one finds that the scattered wave consists of a mixture of electrons
and $u^c$-quarks.

The differential cross section per unit length for the production of
$u$-quark is defined by
\begin{equation}
        \frac{d\sigma}{d\theta} = \lim_{r\rightarrow \infty}
                      \frac{\vec{J}_{sca}^u\cdot \vec{r}}{J_{inc}}
\end{equation}
where $J^i = \bar{\psi}\gamma^i \psi\ $.  Substituting
$\tilde{\psi}_{0\,inc}$ and $\tilde{\psi}_{0\,sca}$
into the currents, one obtains
\begin{equation}
         \frac{d\sigma}{d\theta} =
  \frac{1}{4} \left| f^{\frac{1}{2}}(\theta)
                -f^{-\frac{1}{2}}(\theta) \right|^2\,,
\end{equation}
which can be written out using Eq.~(57) as
\begin{equation}
         \frac{d\sigma}{d\theta} = \frac{1}{2\pi k}
     \left\{ \frac{\cos^4 \frac{\theta}{2}}{\sin^2 \frac{\theta}{2}}
      + \sin^2 \left( \frac{\theta}{2} - 2\delta \right)\right\}\,.
\end{equation}

The calculation above was done in the limit of zero string width.
Now let us examine the string core.  The structure of the string
core is ``encoded'' in the self-adjoint parameter $\delta$ (or
$\mu$, related to $\delta$ by Eq.~(58)), which appears in the
differential cross section above.  In general the self-adjoint
parameter is determined either from physical properties at the
origin or sometimes by symmetry arguments.  Since the string
solutions have already been obtained in the previous section,
we can find $\mu$ by solving Eq.~(50) numerically for the mode
$-1 < n+\lambda_i < 0$, using the realistic form $g(r)/r$ for the
gauge field computed earlier in place of the $1/r$ in Eq.~(50).
As we have shown, $\lambda_i=\pm\frac{1}{2}$ for the fermions that
scatter nontrivially off the $\tau_1$-string.  Thus the special
mode satisfying $-1 < n+\lambda_i < 0$ takes the value $n+\lambda_i
=-\frac{1}{2}$, where $n=-1$ for $\lambda_i=\frac{1}{2}$ and
$n=0$ for $\lambda_i=-\frac{1}{2}$.  Recall that in the calculation
of $g(r)$, the radial distance $r$ was rescaled to the dimensionless
$\sqrt{-v_1} r\ (v_1 < 0)$, where $v_1$ is the quadratic coupling
in the Higgs potential in Eq.~(16).  Rescaling $\chi_2$ and $r$ by
\begin{eqnarray}
	\chi_2 & \rightarrow & i\frac{E+m}{k}\chi_2\,,\nonumber\\
	r & \rightarrow & \sqrt{-v_1} r\,,
\end{eqnarray}
and replacing $\lambda_i$ in Eq.~(50) by $\lambda_i g(r)$,
Eq.~(50) can be rewritten as
\begin{eqnarray}
   \partial_r \chi_1 & = & \frac{g(r)-2}{2r}\chi_1+\beta\chi_2
	\nonumber\\
   \partial_r \chi_2 & = & - \frac{g(r)}{2r}\chi_2
	-\beta\chi_1
\end{eqnarray}
for $\lambda_i=\frac{1}{2}, n=-1$, and
\begin{eqnarray}
	 \partial_r \bar{\chi}_1 & = & -\frac{g(r)}{2r}\bar{\chi}_1
	+ \beta\bar{\chi}_2 \nonumber\\
         \partial_r \bar{\chi}_2 & = & \frac{g(r)-2}{2r}\bar{\chi}_2
	-\beta\bar{\chi}_1
\end{eqnarray}
for $\lambda_i=-\frac{1}{2}, n=0$.  The parameter $\beta$
is defined by
\begin{equation}
	\beta \equiv k/\sqrt{-v_1}\,,
\end{equation}
and the bars over $\chi_1, \chi_2$ are used to distinguish the
solutions of $\lambda_i=-\frac{1}{2}$ from those of $\lambda_i
=\frac{1}{2}$.  Upon closer inspection of the two sets of equations
above, one finds that Eq.~(69) is in fact identical to Eq.~(68) if
$\bar{\chi}_1$ is identified with $\chi_2$ and $\bar{\chi}_2$
with $-\chi_1$.  What about the boundary conditions at the origin?
In Eq.~(49), for $n=-1$, the upper component depends on $\theta$
but the lower component does not, and vice versa for $n=0$.
Therefore $\chi_1$ and $\bar{\chi}_2$ must vanish at the origin for
the solution to be continuous, but $\chi_2$ and $\bar{\chi}_1$ can
be nonzero at $r=0$.  One thus has $\bar{\chi}_1 = \chi_2$
and $\bar{\chi}_2 = -\chi_1$.  Since Eq.~(68) is linear, the
value of $\chi_2(0)$ can be chosen arbitrarily when integrating
the differential equations.

The self-adjoint parameters $\mu$ for $\lambda_i=\frac{1}{2}$ and
$\bar\mu$ for $\lambda_i=-\frac{1}{2}$ are determined by matching
the solutions of Eq.~(68) to the asymptotic expression in Eq.~(52)
at some radius $r$.  For $n+\lambda_i = -\frac{1}{2}$, the Bessel
functions in Eq.~(52) are simply $J_{\pm\frac{1}{2}}$, which have
the analytic forms
\begin{equation}
	J_{\frac{1}{2}}(x)=\sqrt{\frac{2}{\pi x}} \sin x\,,\quad
	J_{-\frac{1}{2}}(x)=\sqrt{\frac{2}{\pi x}} \cos x\,.
\end{equation}
Then Eq.~(52) leads to the simple expression for $\mu$ and
$\bar\mu$:
\begin{eqnarray}
	\frac{\chi_1}{\chi_2} &=& \tan(\mu + \beta r) \,,\nonumber\\
	\frac{\bar{\chi}_1}{\bar{\chi}_2} &=&
	\tan(\bar\mu + \beta r) \,,
\end{eqnarray}
which can be inverted to give $\mu$ and $\bar\mu$ at a given $r$,
using $\chi_1$ and $\chi_2$ computed from Eq.~(68).
Using Eq.~(72) and trigonometric identities, one finds
\begin{equation}
	\bar\mu = \mu + \frac{\pi}{2} \,.
\end{equation}

Note that the solutions depend on $\beta$ which appears
in Eq.~(68), and the quartic couplings $\lambda_2, \lambda_3$ in
the Higgs potential.  The parameter $\beta$ defined in Eq.~(70)
measures the ratio of the incident fermion momentum $k$ to
the Higgs mass parameter $\sqrt{-v_1}$, which is of the order of
GUT energy scale.  To put it another way, $\beta$ measures
the string width relative to the wavelength of the incident fermion.
In Fig.~5, we set $\beta = 1$ and plot $\mu$ computed from Eq.~(72)
at a given $r$ for three sets of $\lambda_2$ and $\lambda_3$.
The true value of $\mu$ is given by the limit $r \rightarrow
\infty$.  In Fig.~6, we choose the same set of parameter as in
Figs. 1-3: $\lambda_2 = 0.132$ and $\lambda_3 = 10.25$; $\mu$
is shown for five values of $\beta$ ranging from 0.1 to 2.0.
One can see that as $\beta$ decreases, {\it i.e.} when the
wavelength of the fermion becomes large compared to the string
width, $\mu$ decreases.

\section{CONCLUSIONS}
We constructed two types of strings, string-$\tau_{\rm all}$
and string-$\tau_1$, in the SO(10) grand unified theory.
They are topologically equivalent but dynamically different
strings, produced during the phase transition $\hbox{Spin(10)}
\rightarrow\hbox{SU(5)}\times{\cal Z}_2$ in the early universe.
String-$\tau_{\rm all}$ is effectively Abelian, and can catalyze
baryon number violation with a strong cross section via grand-unified
processes inside the string.  It has been the subject of study in
several recent papers.  The richer Higgs structure of
string-$\tau_1$, on the other hand, has been shown in this paper
to induce baryon catalysis by mixing components in the
fermion multiplet, turning leptons into quarks as they travel
around the string.  The underlying B-violating mechanism is
the ``twisting'' of the scalar field, which leads to different
unbroken SU(5) subgroups around the string.  This mechanism is
distinct from the grand-unified processes which can only occur
inside the string core where the GUT symmetry is restored.

The corresponding string solutions have been calculated numerically
with both the ``shooting'' and the relaxation methods.  The energy
of both strings was computed.  With an additional upper bound
argument, we found string-$\tau_1$ to have lower energy than
string-$\tau_{\rm all}$ in a wide range of parameters:
$10^{-4} < \lambda_2 < 2.5\times 10^3$ and all $\lambda_3$.  The
ratio of the upper bound on $\tau_1$ energy to the $\tau_{\rm all}$
energy increases as $\lambda_2$ decreases, and possibly approaches
one from below as $\lambda_2 \rightarrow 0$.  Scattering of fermions
in the fields of string-$\tau_1$ has also been analyzed, and the
B-violating cross section is given by Eq.~(66).  We conclude that
string-$\tau_1$ is more stable than string-$\tau_{\rm all}$, and
can catalyze baryon decay with strong cross sections via the
interesting mechanism of Higgs field twisting.

\nonum
\section{ACKNOWLEDGMENTS}
I wish to thank Alan Guth for many valuable suggestions on
this work and a critical reading of the manuscript.  I am also
grateful for advice from Ed Bertschinger, Robert Brandenberger,
Jeffrey Goldstone, Roman Jackiw and Leandros Perivolaropoulos,
and assistance from Roger Gilson.

\nonum
\section{APPENDIX}

The generators of SO(2n) in the spinor representation can be
constructed from a set of $2^n \times 2^n$ hermitian matrices
$\Gamma_a^{(n)}, a=1, \ldots ,2n\,$, which satisfy the Clifford
algebra
\begin{equation}
        \{\Gamma_a^{(n)},\Gamma_b^{(n)}\} = 2\delta_{ab}\,.
	\eqnum{A.1}
\end{equation}
Starting with the two Pauli matrices for $n=1$
\begin{equation}
        \Gamma_1^{(1)} = \left( \begin{array}{cc}
                                0\ & 1\ \\
                                1\ & 0\
                        \end{array} \right) \,, \quad
        \Gamma_2^{(1)} = \left( \begin{array}{cc}
                                0 & -i \\
                                i & 0
                        \end{array} \right) \,, \eqnum{A.2}
\end{equation}
one can iteratively build the higher-dimensional
$\Gamma^{(n+1)}_a\ $ from the $\Gamma^{(n)}_a\ $ by
\begin{eqnarray}
  \Gamma_a^{(n+1)} &=& \left( \begin{array}{cc}
                        \Gamma_a^{(n)} & 0 \\
                         0  & -\Gamma_a^{(n)}
                        \end{array} \right)\,,\ a=1,\ldots ,2n
                                                \nonumber\\
  \Gamma_{2n+1}^{(n+1)} &=& \left( \begin{array}{cc}
                         0\ \  & 1 \\
                         1\ \  & 0
                        \end{array} \right)\,, \nonumber\\
  \Gamma_{2n+2}^{(n+1)} &=& \left( \begin{array}{cc}
                         0\  & -i \\
                         i\  & 0
                        \end{array} \right)\,. 	\eqnum{A.3}
\end{eqnarray}
One can check that these $\Gamma$ matrices satisfy the Clifford
algebra.  The $\frac{2n(2n-1)}{2}$ generators of SO(2n) are
constructed by
\begin{equation}
    M_{ab} = \frac{1}{4i} [\Gamma_a,\Gamma_b]\,,\ \ a,b=1,
	\ldots ,2n 	\eqnum{A.4}
\label{eq:clifford}
\end{equation}
where $M_{ab}$ satisfy the SO(2n) commutation relations
\FL
\begin{equation}
   [M_{ab},M_{cd}]=-i(\delta_{bc} M_{ad}+\delta_{ad} M_{bc}
        -\delta_{ac} M_{bd}-\delta_{bd} M_{ac})\,. \eqnum{A.5}
\end{equation}

Thus far, we have used the explicit matrix notation to
construct $\Gamma$ and $M$.  For convenience, however, we
will use an alternative notation in which each of the
$2^n \times 2^n$ matrices is written as a tensor product
of $n$ independent Pauli matrices, each acting on a different
two-dimensional space.  We choose the convention that the first
matrix from the right in the tensor product acts on the largest
2$\times$2 block in the matrix notation, while the second from
the right acts on the next, and so on, with the matrix on the left
acting on the smallest 2$\times$2 block.  In this notation,
the 10 $\Gamma$'s of SO(10) given by Eq.~(A.3) become
\begin{eqnarray}
 \Gamma_1 &= \sigma_1 \sigma_3 \sigma_3 \sigma_3 \sigma_3\,,\ &
 \Gamma_2 = \sigma_2 \sigma_3 \sigma_3 \sigma_3 \sigma_3\,,
	\nonumber\\
 \Gamma_3 &= I\ \sigma_1 \sigma_3 \sigma_3 \sigma_3\,,&
 \Gamma_4 = I\ \sigma_2 \sigma_3 \sigma_3 \sigma_3\,,\nonumber\\
 \Gamma_5 &= I\ I\ \sigma_1 \sigma_3 \sigma_3\,,&
 \Gamma_6 = I\ I\ \sigma_2 \sigma_3 \sigma_3\,,\nonumber\\
 \Gamma_7 &= I\ I\ I\ \sigma_1 \sigma_3\,,&
 \Gamma_8 = I\ I\ I\ \sigma_2 \sigma_3\,,\nonumber\\
 \Gamma_9 &= I\ I\ I\ I\ \sigma_1\,,&
 \Gamma_{10} = I\ I\ I\ I\ \sigma_2 \,,
 \eqnum{A.6}
\end{eqnarray}
and the 45 generators $M$ can be found accordingly.
Furthermore, one can write down the five diagonal $M$'s
that generate the Cartan sub-algebra:
\begin{eqnarray}
        M_{12} &=& \frac{1}{2}\ \sigma_3 I I I I\,, \nonumber\\
        M_{34} &=& \frac{1}{2}\ I \sigma_3  I I I\,, \nonumber\\
        M_{56} &=& \frac{1}{2}\ I I\sigma_3 I I\,, \nonumber\\
        M_{78} &=& \frac{1}{2}\ I I I\sigma_3  I\,, \nonumber\\
        M_{9\,10} &=& \frac{1}{2}\ I I I I\sigma_3 \,.
	 \eqnum{A.7}
\end{eqnarray}

The eigenvalues of the five generators above can be used
to label the states in the spinor representation.  Let
$\frac{1}{2}\epsilon_1, \ldots , \frac{1}{2}\epsilon_5$
be the eigenvalues of $M_{12}, \ldots ,M_{9\,10}$ respectively
with $\epsilon_i = +1$ or $-1$, and denote the states by
\begin{equation}
         |\,\epsilon_1 \epsilon_2 \epsilon_3 \epsilon_4 \epsilon_5
              \,\rangle\,.  \eqnum{A.8}
\end{equation}
This 32-dimensional representation is reducible to two 16-dimensional
irreducible representations because there exists a chirality operator
\begin{eqnarray}
    \chi &\equiv & (-i)^5 \Gamma_1 \Gamma_2 \ldots \Gamma_{10}
                        \nonumber\\
         & =     &  \sigma_3\sigma_3\sigma_3\sigma_3\sigma_3\,,
	 \eqnum{A.9}
\end{eqnarray}
which satisfies the commutation relations
\begin{equation}
        \{\,\chi, \Gamma_i\,\} = 0\,,\quad [\,\chi, M_{ab}\,]=0\,.
	 \eqnum{A.10}
\end{equation}
Moreover,
\begin{equation}
    \chi |\,\epsilon_1 \epsilon_2 \epsilon_3 \epsilon_4 \epsilon_5
        \,\rangle       \ = \prod_{i} \epsilon_i
    |\,\epsilon_1 \epsilon_2 \epsilon_3 \epsilon_4
        \epsilon_5\,\rangle\,,   \eqnum{A.11}
\end{equation}
where the eigenvalue $\prod_{i} \epsilon_i$ is $+1$ or $-1$
depending on whether the number of spins that are down
$(\epsilon_i=-1)$ is even or odd.

We assign the sixteen left-handed fermions to the states
of positive chirality, {\it i.e. } states with even number of
$\epsilon_i = -1$.  The explicit identification of states
to fermions can be achieved by first breaking the SO(10)
10$\times$10 representation into an upper 6$\times$6 and
a lower 4$\times$4 blocks for the subgroups SO(6) and SO(4),
and then embedding SU(3) in SO(6) and SU(2) in SO(4).
The generators for SO(4) are $M_{ab}, a, b = 7,8,9,10$,
and with the choice \cite{spin}
\begin{equation}
      \tau_i=\frac{1}{2} \epsilon_{ijk} M_{jk} - M_{i\,10}\,,
        \ \ i,j,k = 7,8,9    \eqnum{A.12}
\end{equation}
for the generators of SU(2), one can easily verify that
the last two spins in $|\,\epsilon_1 \epsilon_2 \epsilon_3
\epsilon_4 \epsilon_5\,\rangle$ label the SU(2) states with
$|+ -\,\rangle, |- +\,\rangle$ labeling the doublets and
$|+ +\,\rangle, |- -\,\rangle$ the singlets.
Similarly, the first three spins in
$|\epsilon_1 \epsilon_2 \epsilon_3 \epsilon_4 \epsilon_5 \rangle$
label the SU(3) states with $|+ + +\,\rangle, |- - - \,\rangle$
labeling the singlets, and $|+ + -\,\rangle, |- + + \,\rangle$
with their permutations labeling the SU(3) triplets.  One also
needs the charge operator $Q$ to make the assignment unique.
In SU(5), $Q = diag(1/3,1/3,1/3,0,-1)$, which takes the form
\begin{equation}
        Q = \frac{1}{3} ( M_{12} + M_{34} + M_{56} ) - M_{9\,10}\,.
	 \eqnum{A.13}
\end{equation}
In the SO(10) spinor representation,
\begin{equation}
     Q |\,\epsilon_1...\epsilon_5\,\rangle
    = \left\{ \frac{1}{6} (\epsilon_1 + \epsilon_2 + \epsilon_3)
    - \frac{\epsilon_5}{2} \right\} |\,\epsilon_1 \ldots \epsilon_5
        \,\rangle\,.    \eqnum{A.14}
\end{equation}
Putting all the above together one obtains
\begin{eqnarray}
     |+ + + + +\,\rangle &= \nu^c\,,\ & |+ + + - -\,\rangle = e^+
              \nonumber\\
     |- - + + +\,\rangle &= u^c_1\,,\ & |- - + - -\,\rangle = d^c_1
                \nonumber\\
     |- + - + +\,\rangle &= u^c_2\,,\ & |- + - - -\,\rangle = d^c_2
                \nonumber\\
     |+ - - + +\,\rangle &= u^c_3\,,\ & |+ - - - -\,\rangle = d^c_3
                \nonumber\\
     |- - - + -\,\rangle &= \nu\,,\ & |- - - - +\,\rangle = e^-
                \nonumber\\
     |+ + - + -\,\rangle &= u_1\,,\ & |+ + - - +\,\rangle = d_1
               \nonumber\\
     |+ - + + -\,\rangle &= u_2\,,\ & |+ - + - +\,\rangle = d_2
	    \nonumber\\
     |- + + + -\,\rangle &= u_3\,,\ & |- + + - +\,\rangle = d_3\,.
\eqnum{A.15}
\end{eqnarray}
Since we already know how to express the generators $M_{ab}$
as matrices, we can write the states as a single 32-dimensional
column vector which is projected into two 16-dimensional vectors
of positive and negative chirality by the operator
$P_\pm \equiv \frac{1}{2} (1\pm\chi)$.  We find
\FL
\begin{equation}
	\psi = (\nu^c\ u^c_1\ u^c_2\ u^c_3\ d_3\ d_2\ d_1\ e^-
	     \ u_3\ u_2\ u_1\ \nu\ e^+\ d^c_1\ d^c_2\
	     d^c_3)_L\,.   \eqnum{A.16}
\end{equation}

In this paper, we studied two types of strings:
string-$\tau_{\rm all}$, where $\tau_{\rm all}$ is given by
Eq.~(10), and string-$\tau_1$, where $\tau_1$ can be any of
the generators in Eq.~(12).  It is easy to see that in terms
of $M_{ab}, \tau_{\rm all}$ is written as
\begin{equation}
   \tau_{\rm all} = \frac{1}{5}
	(M_{12}+M_{34}+M_{56}+M_{78}+M_{9\,10})\,,  \eqnum{A.17}
\end{equation}
and $|\,\epsilon_1  \ldots  \epsilon_5\,\rangle$ is an eigenstate of
$\tau_{\rm all}$ with eigenvalue $\frac{1}{10} \sum_i \epsilon_i\,.$
For the left-handed fermions above, $\frac{1}{10} \sum_i \epsilon_i
= \frac{1}{2}$ for $\nu^c$, $\frac{1}{10}$ for $e^+, u, d, u^c$,
and $-\frac{3}{10}$ for $\nu, e^-, d^c$.

To study how $\tau_1$ act on the fermions, we write $\tau_{1+}$
and $\tau_{1-}$ defined in Eq.~(23) as a product of five Pauli
matrices using Eqs.~(A.4) and (A.6), and then replace the
matrices $\sigma_1$ and $\sigma_2$ by the usual raising and
lowering operators $\sigma_\pm=\frac{1}{2} (\sigma_1 \pm i\sigma_2)$.
One obtains
\begin{eqnarray}
  \tau_{1+} & = & \frac{1}{2} (\tau^{2\alpha-1, 2\beta}
                        + \tau^{2\alpha, 2\beta-1}) \nonumber\\
            & = & I \ldots I \sigma_+ \sigma_3 \ldots \sigma_3
		\sigma_+ I \ldots I \nonumber\\
             &&   + I \ldots I \sigma_- \sigma_3 \ldots \sigma_3
		\sigma_- I \ldots I   \eqnum{A.18}
\end{eqnarray}
and
\begin{eqnarray}
  \tau_{1-} & = & \frac{1}{2} (\tau^{2\alpha-1, 2\beta-1}
                        - \tau^{2\alpha, 2\beta}) \nonumber\\
            & = & I \ldots I \sigma_+ \sigma_3 \ldots \sigma_3
		\sigma_- I \ldots I \nonumber\\
            && - I \ldots I \sigma_- \sigma_3 \ldots \sigma_3
		\sigma_+ I \ldots I   \eqnum{A.19}
\end{eqnarray}
where $\alpha,\beta=1, \ldots 5, \alpha < \beta$, and
the two $\sigma_\pm$ matrices in each term occupy the
$\alpha$th and $\beta$th positions from the left.  Now one
can read off from the list of fermions above which particles
are mixed by a given $\tau_1$.  For generators of the form
$\tau_{1+}$, one immediately finds that except for the case
$\alpha=4, \beta=5$, all mix leptons with quarks; when
$\alpha=4, \beta=5$, the generator mixes $(e^+, \nu^c),
(u_1^c, d_1^c), (u_2^c, d_2^c),$ and $(u_3^c, d_3^c)$.
For generators of the form $\tau_{1-}$, leptons are mixed
with quarks when $\alpha$ = 1, 2, or 3 and $\beta$ = 4 or 5.

\newpage
\figure{The solution of string-$\tau_1$, $g(r), f_o(r), f_e(r)$,
as a function of dimensionless $r$ for the case
$\lambda_2=0.132, \lambda_3=10.25$.  The function $g(r)$
represents the spatial dependence of the gauge field, and
$f_o(r), f_e(r)$ represent that of the Higgs field. \label{fig1}}
\figure{The solution of string-$\tau_{all}$, $g(r), f(r)$,
as a function of dimensionless $r$ for the same case as in
Fig.~\ref{fig1}.  Here $g(r)$ represents the spatial dependence of
the gauge field and $f(r)$ that of the Higgs field. \label{fig2}}
\figure{The radial energy density $2\pi r {\cal H}(r)$
(in units of $v_1^2/v_2$) of string-$\tau_1$ and $\tau_{\rm all}$,
computed from the solutions in Figs.~1 and 2. \label{fig3}}
\figure{The ratio of the upper bound on $\tau_1$ energy,
$E_1(f_o=f_e)$, over the $\tau_{\rm all}$ energy, $E_{\rm all}$,
as a function of $\lambda_2$.  $E_1(f_o=f_e)$ is calculated by
setting $f_o=f_e$ in the Lagrangian. \label{fig4}}
\figure{The self-adjoint parameter $\mu$ computed from Eq.~(72)
at a given $r$, for three sets of $(\lambda_2,\lambda_3)$:
(0.132, 10.25), (0.264, 20.50) and (0.528, 41.0), where $\beta=1$.
The true value of $\mu$ is given in the limit $r \rightarrow
\infty$. \label{fig5}}
\figure{The self-adjoint parameter $\mu$ computed from Eq.~(72) at
a given $r$ for different ratios of $\beta$, where $\lambda_2 =
0.132, \lambda_3 = 10.25$. \label{fig6}}

\end{document}